\documentclass[twocolumn,showpacs,preprintnumbers,amsmath,
               nofootinbib,aps,superscriptaddress]{revtex4}

\usepackage{graphicx}
\usepackage{dcolumn}
\usepackage{bm}

\newcommand*{\bea}{\begin{eqnarray}}
\newcommand*{\eea}{\end{eqnarray}}
\newcommand*{\be}{\begin{equation}}
\newcommand*{\ee}{\end{equation}}

\newcommand*{\pref}[1]{(\ref{#1})}

\newcommand*{\tl}{\mathrm{tl}}

\begin{document}

\preprint{}

\title{Three-point vertices in Landau-gauge Yang-Mills theory}

\author{Attilio Cucchieri}\email{attilio@ifsc.usp.br}
\affiliation{Instituto de F\'\i sica de S\~ao Carlos, Universidade de S\~ao Paulo, \\
             Caixa Postal 369, 13560-970 S\~ao Carlos, SP, Brazil}

\author{Axel Maas}\email{axel.maas@uni-graz.at}
\affiliation{Department of Complex Physical Systems, Institute of Physics, \\
             Slovak Academy of Sciences, D\'{u}bravsk\'{a} cesta 9, SK-845 11 Bratislava,
             Slovakia}
\affiliation{Department of Theoretical Physics, Institute of Physics,\\
             Karl-Franzens University Graz, Universit\"atsplatz 5, A-8010 Graz,
             Austria}

\author{Tereza Mendes}\email{mendes@ifsc.usp.br}
\affiliation{Instituto de F\'\i sica de S\~ao Carlos, Universidade de S\~ao Paulo, \\
             Caixa Postal 369, 13560-970 S\~ao Carlos, SP, Brazil}
\affiliation{DESY--Zeuthen, Platanenallee 6, 15738 Zeuthen, Germany}

\date{\today}

\begin{abstract}
Vertices are of central importance for constructing QCD bound states out of the
individual constituents of the theory, i.e.\ quarks and gluons. In particular, the determination
of three-point vertices is crucial in non-perturbative investigations of QCD. We use
numerical simulations of lattice gauge theory to obtain results for the 3-point vertices
in Landau-gauge SU(2) Yang-Mills theory in three and four space-time dimensions for various
kinematic configurations. In all cases considered, the ghost-gluon vertex is found
to be essentially tree-level-like, while the three-gluon vertex is suppressed at
intermediate momenta. For the smallest physical momenta, reachable only in three dimensions,
we find that some of the three-gluon-vertex tensor structures change sign.
\end{abstract}

\maketitle


\section{Introduction}

Vertices describe the basic interactions between the elementary degrees
of freedom of QCD and are thus of central importance for the understanding
of non-trivial phenomena in the physics of strong interactions. The properties
of vertices, in particular in the momentum regime of the average constituent
momentum, are crucial for the formation of bound states. Furthermore, the
far-infrared behavior of vertices should be connected to the confining properties
of the theory, since confinement necessarily originates in the interaction of the
fields. Quite clearly, a determination of these vertices is an important step in
any understanding of the non-perturbative regime of QCD. Finally, in QCD the
vertices are also important for the breaking of chiral symmetry, and thus are
a central ingredient in the understanding of hadron physics.

In general, the vertices are gauge-dependent quantities. This means that one
must understand how gauge-invariant objects, such as hadrons, are constructed
from gauge-variant objects, i.e.\ quarks and gluons. This question has
been widely studied by analytical methods, using in particular Dyson-Schwinger
equations \cite{Alkofer:2000wg,roberts,Fischer:2006ub}. Most of these studies
were carried out either in Landau or in Coulomb gauge and specific models were
used for the vertices. Here we consider the (minimal) Landau gauge
in order to add results for these vertices from lattice gauge theory. We
concentrate only on the vertices in SU(2) Yang-Mills theory. A thorough
study of the SU(3) case, as well as of the quark-gluon vertex \cite{Alkofer:2006gz,
Llanes-Estrada:2004jz,qvlat}, is the next logical step. Let us stress, however, that
recent lattice studies (in the quenched case) \cite{SU2SU3}
have provided support to the analytic prediction \cite{Alkofer:2000wg}
of identical infrared behavior
for Landau-gauge gluon and ghost propagators in SU(2) and SU(3) gauge theory
in any dimension.

We note that predictions for the infrared behavior of all Green's functions
in Landau gauge have been made in four \cite{Alkofer:2006gz,Lerche:2002ep,Pawlowski:2003hq,
Alkofer:2004it,Alkofer:2008jy,Kellermann:2008iw}
and also in lower dimensions \cite{Huber:2007kc}. These predictions are claimed
to be unique under certain assumptions \cite{Fischer:2006vf}.
Furthermore, they yield an infrared enhanced ghost propagator which is accompanied
by an infrared finite or vanishing gluon propagator. In particular,
an infrared tree-level-like ghost-gluon vertex leads to an infrared vanishing gluon
propagator \cite{Lerche:2002ep}.

The analytic results have been extensively tested
in lattice gauge theory studies \cite{Cucchieri:2007ta,4d,Oliveira:2007dy,gccvertex,
Cucchieri:2006tf,Cucchieri:2003di}
and were confirmed in two dimensions \cite{Maas:2007uv}. In higher
dimensions a major obstacle are finite-volume effects \cite{Cucchieri:2007ta,
Cucchieri:2003di,Cucchieri:2006xi,Maas:2007uv,Fischer:2007pf,cucchieril7}. Nevertheless,
it has been recently shown \cite{Cucchieri:2007rg} that one can control the
infinite-volume extrapolation of the data for the gluon propagator $D(p)$
by considering rigorous lower and
upper bounds, expressed in terms of the momentum-space gluon field. As a result, it was
found that the Landau-gauge gluon propagator at zero momentum $D(0)$ is finite and
nonzero in 3d and in 4d, while $D(0) = 0$ in 2d, in agreement with Ref.\ \cite{Maas:2007uv}.
At the same time, the infrared enhancement of the ghost propagator seems to disappear
when large lattice volumes are considered \cite{Oliveira:2007dy,cucchieril7}.
However, in Refs.\ \cite{Bogolubsky} it has been claimed that the analysis of
Gribov-Singer-copies effects \cite{Gribov} may modify these results,
even though these effects seem to diminish, albeit slowly, with increasing
volume. In any case, a better understanding might be obtained by considering
upper and lower bounds also for the ghost propagator \cite{limghost}.

Let us remark that in Ref.\ \cite{Dudal:2007cw} it was shown that one can also
obtain a finite $D(0)$ gluon propagator and a tree-level-like ghost propagator at small
momenta using the Gribov-Zwanziger approach. Similar results, using numerical solutions
of the Dyson-Schwinger equations, have been claimed in \cite{otherDSE}
\footnote{However, see the remarks in \cite{Fischer:2006ub} on these solutions.}.

Here we consider the simplest vertices, i.e.\ the three-point vertices. Two of these
exist in Landau gauge: the ghost-gluon vertex and the three-gluon vertex. We present
results in four dimensions in Section \ref{s4d}. As we will show, we find an essentially
tree-level-like ghost-gluon vertex and a three-gluon vertex that is suppressed at
intermediate and small momenta. We note, however, that in four dimensions most 
of our lattice volumes are rather small and the statistic for the largest 
volume is low, i.e.\ for the three-gluon vertex we cannot
really probe the asymptotic infrared limit. Therefore, we also
present results in three dimensions, in Section \ref{s3d}. In this case we find a
qualitative change for the three-gluon vertex. More precisely, this vertex
shows a sign change and an enhancement (in absolute value) in the far infrared regime.

Technical details of the lattice calculations are given in Appendix \ref{atech}. In
particular, the lattice parameters employed are reported in Table \ref{conf3}. For the
sake of completeness, the data for the propagators are presented in Appendix \ref{aprop}.

This work extends previous studies in three and in four dimensions \cite{Cucchieri:2006tf,
gccvertex,gggvertex,Alles:1996ka} including additional kinematic configurations and larger
lattice volumes. Preliminary results have been reported in \cite{Maas:2006qw}.


\section{Vertices in four dimensions}\label{s4d}

\begin{figure*}
\includegraphics[width=\linewidth]{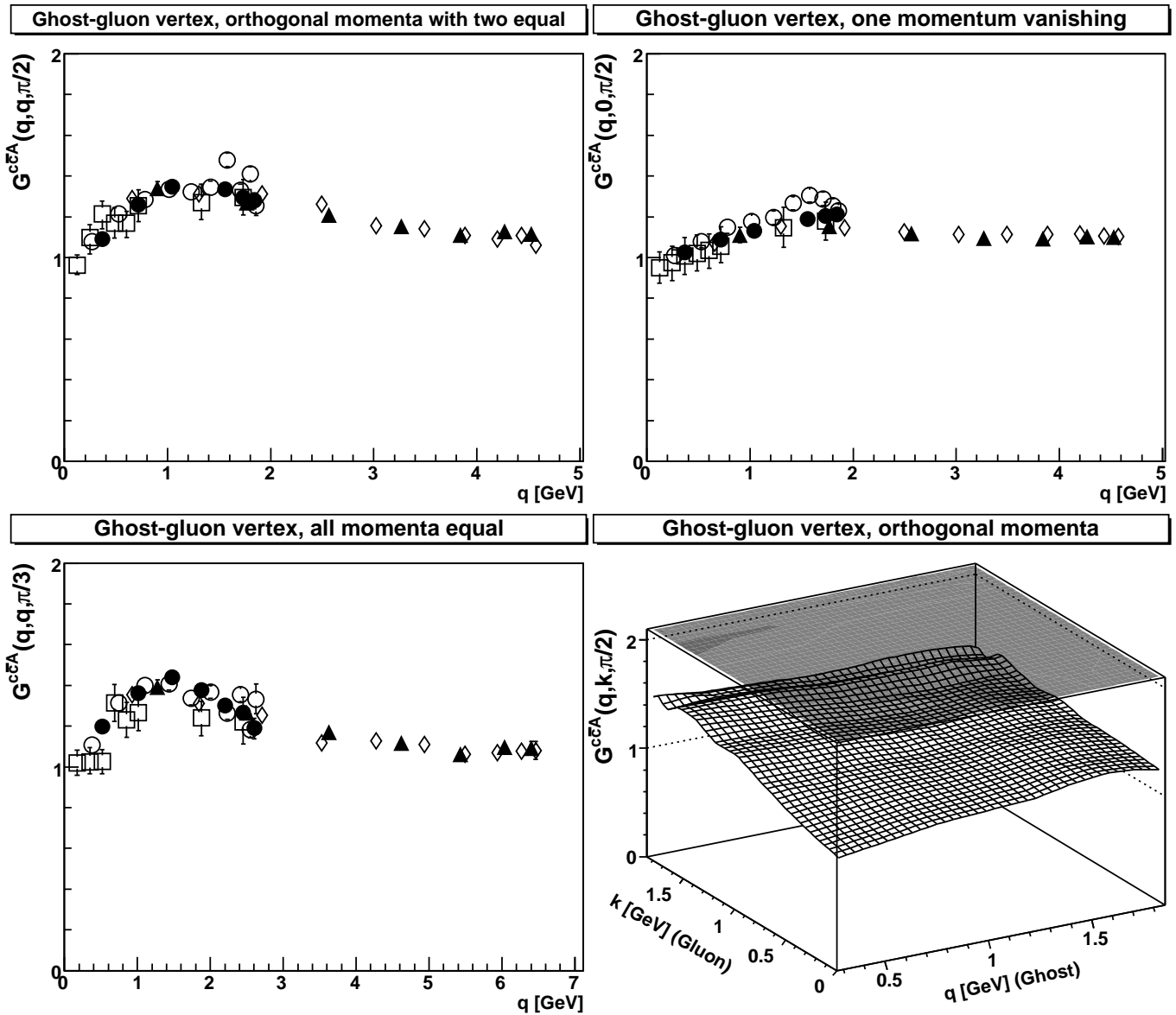}
\caption{\label{f4dggv}
The ghost-gluon vertex in four dimensions. The bottom-left panel shows the
so-called equal configuration. The bottom-right panel shows the orthogonal
configuration. The top panels give cuts through the orthogonal plane. The
top-left panel shows the case of ghost and gluon momentum with equal magnitude.
The top-right panel shows the case with a vanishing gluon momentum.
Full triangles are from a $16^4$ lattice at $\beta=2.5$,
open diamonds from a $22^4$ lattice at $\beta=2.5$,
full circles from a $16^4$ lattice at $\beta=2.2$,
open circles from a $22^4$ lattice at $\beta=2.2$
and open squares from a $48^4$ lattice at $\beta=2.2$.
The bottom-right panel shows the data only
for the $22^4$ lattice at $\beta=2.2$. All data are in physical units.}
\end{figure*}

\begin{figure*}
\includegraphics[width=\linewidth]{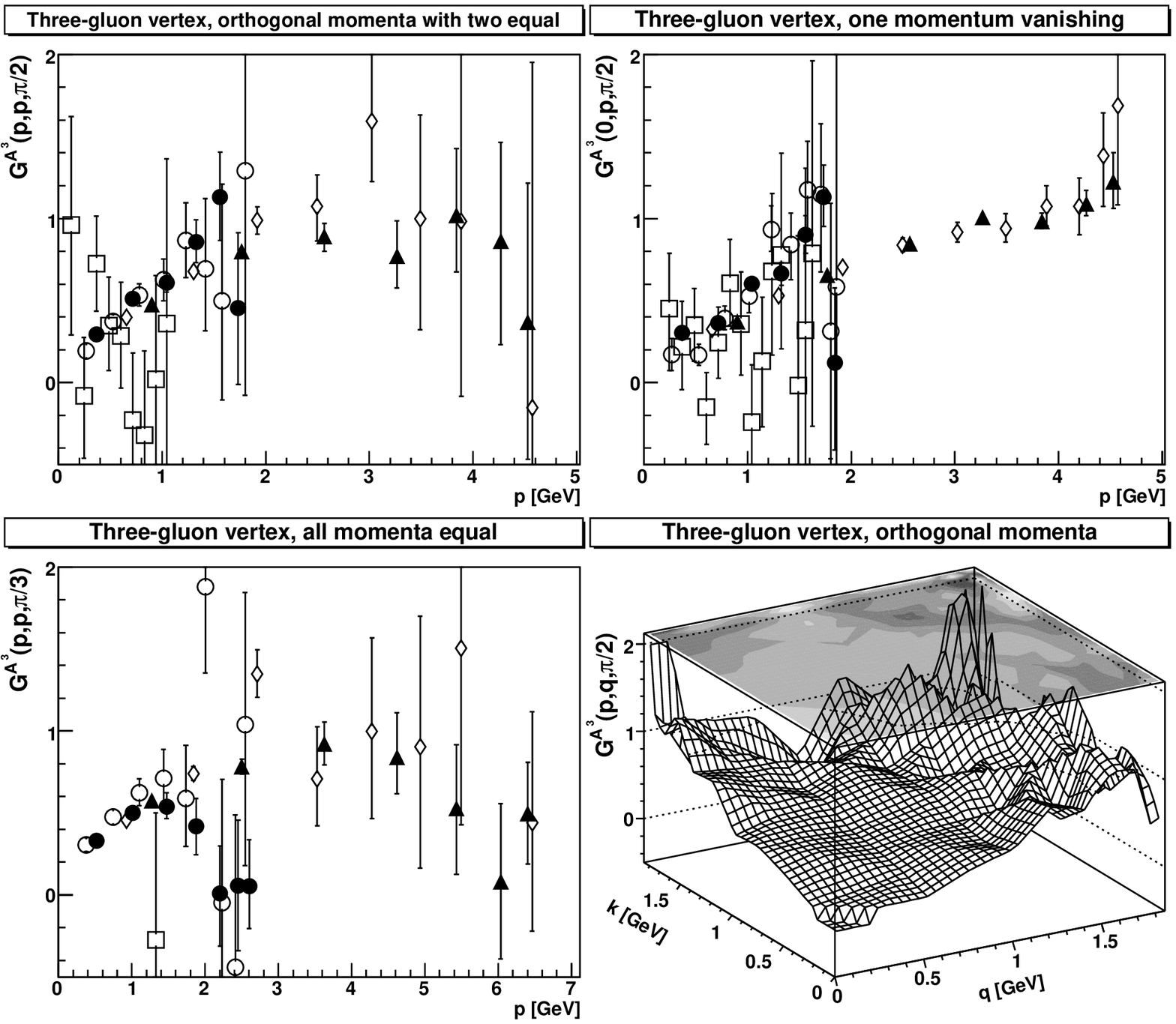}
\caption{\label{f4dg3v}
Same as in Figure \ref{f4dggv} but the data now refer to the three-gluon vertex
in four dimensions. Note that, due to the relatively low statistics in the $48^4$
case, plot points with an absolute error large than 1.5
have been dropped for this system.}
\end{figure*}

In lattice calculations, one can easily evaluate the full Green's functions, while the vertex
functions cannot be computed directly. Let us recall that the ghost-gluon vertex has only one
(color-antisymmetric) tensor structure in its full Green's function. On the contrary, the
three-gluon vertex has a much richer structure. It is possible to determine the contribution
of each tensor structure by considering the appropriate projection of the full Green's function.
Here only one such tensor structure will be investigated, the one given by the
projection\footnote{The 
collinear singularities discussed in \cite{Alkofer:2008jy} are not affecting the tensor structure
considered here. We are grateful to Markus Huber and Kai Schwenzer for providing this
information.}
\be
G = \frac{\Gamma^\tl_{abc} D_{ad} D_{be} D_{cf} \Gamma_{def}}{
          \Gamma^\tl_{abc} D_{ad} D_{be} D_{cf} \Gamma^\tl_{def}} ,
\label{proj}
\ee
where $\Gamma$ denotes the vertices, $D$ are (gluon or ghost) propagators and the indices
are multi-indices encompassing field-type, Lorentz and color indices. The superscript $\tl$
indicates tree-level quantities. Note that this quantity is dimensionless.

Clearly, for the ghost-gluon vertex the above tensor structure \pref{proj} reduces to the
single tensor structure characterizing its full Green's function. On the contrary, in the
case of the three-gluon vertex, it is a linear combination of the transverse vertex tensor
structures in the conventional separation scheme discussed in Ref.\ \cite{Ball:1980ax}.
Note that the normalization in \pref{proj} is chosen so as to absorb trivial kinematic
factors, yielding $G$ equal to 1 if the full and the tree-level vertices coincide. A more
detailed discussion of the quantity \pref{proj} can be found in \cite{Cucchieri:2006tf}.

One should also recall that three-point vertices depend on two independent external momenta.
Using translational and rotational invariance, this dependence can be reduced to three kinematic quantities.
These will be chosen as the magnitude of two of the external momenta
and the angle between them. In the case of the three-gluon vertex, due to bosonic symmetry, it is
irrelevant which of the momenta of the three external lines are chosen as independent. For the
ghost-gluon vertex, we consider the gluon and the ghost momenta. Note that, since Landau gauge
is ghost-anti-ghost symmetric \cite{Alkofer:2000wg}, the ghost and anti-ghost momenta can
be exchanged without modifying the result.

Of course, in lattice calculations, the kinematic variables have to be compatible with the
(hyper-cubic) symmetry of the lattice. As discussed in Ref.\ \cite{Cucchieri:2006tf}, we
consider two specific kinematic configurations, denoted respectively as {\em orthogonal} and
{\em equal}. In the first case the two external momenta are chosen orthogonal
to each other, i.e.\ the angle in between is equal to 90 degrees. In the equal case the two
momenta have equal magnitude and the angle is 60 degrees. The first case allows one to reach
the smallest possible non-zero momentum on a given lattice. The second case reduces
the problem to a one-scale problem, which is attractive in studies using functional methods
in the far infrared \cite{Alkofer:2006gz,Alkofer:2004it,Fischer:2006vf,Huber:2007kc}. Indeed,
in that case it is predicted that all $n$-point vertices behave as power laws in
this single external scale \cite{Alkofer:2006gz,Alkofer:2004it,Fischer:2006vf}. In addition,
if the exponent of the power law of one of the vertices is known, the others are all fixed
\cite{Fischer:2006vf}.

The remaining technical details for the determination of vertices can be found in
\cite{Cucchieri:2006tf} and in Appendix \ref{atech}. A list of the studied systems
is given in Table \ref{conf3} in Appendix \ref{atech}.

Our results for the ghost-gluon vertex $G^{c \bar{c} A}$ are shown in Figure \ref{f4dggv}.
For all momentum configurations, the vertex is essentially flat, except for a shallow maximum
at about 1 GeV. Therefore, the ghost-gluon vertex is essentially unmodified compared to
its tree-level version. In particular, this is the case both for the vertex measured in
the equal (symmetric) or in the orthogonal momentum configuration. This result is in
agreement with previous studies in 4d \cite{gccvertex} and with data obtained in
lower dimensional systems (see Section \ref{s3d} and  Refs.\ \cite{Maas:2007uv,Cucchieri:2006tf}).
Furthermore, it is in agreement with results from DSE calculations \cite{Schleifenbaum:2004id}.
As noted above, this result can also be used as an input to solve the
complete tower of functional equations for the Yang-Mills Green's functions.

The results for the three-gluon vertex $G^{A^3}$ are shown in Figure \ref{f4dg3v}. As has
been previously observed \cite{Maas:2006qw}, the vertex is suppressed at mid-momentum
compared to a bare one, independently of the momentum configuration considered.
However, even on the largest lattice it is not clear whether (or not) the vertex becomes
negative at small momenta, or shows a divergence towards zero momentum. Indication of such
a divergence was found in lower dimensions \cite{Maas:2007uv,Cucchieri:2006tf}.
Note that, as in lower dimensions \cite{Cucchieri:2006tf}, there is also a clear increase in the
statistical noise for large lattice momenta. A possible solution to this problem, in order
to obtain a good signal-to-noise ratio also at large physical momenta, would be of course
to simulate at larger values of $\beta$. Unfortunately, the three-gluon vertex is in general very
noisy, and thus the results for the expensive case of the $48^4$ lattice are affected by very
strong statistical fluctuations. Nonetheless, the data points with acceptable errors confirm
the trend seen on the smaller lattices.

As said above, it is difficult to compare the result for the three-gluon vertex $G^{A^3}$ with
results obtained using functional methods, since one needs a sufficiently large statistic
for large lattice volumes, in order to probe the true infrared limit. However, the mid-momentum
suppression observed may already be relevant for phenomenological applications and for
the determination of the quark-gluon vertex using functional methods \cite{Llanes-Estrada:2004jz}.
This suppression is also certainly useful in understanding the discrepancies obtained for 
the propagators when comparing results from functional methods to lattice studies
\cite{Alkofer:2000wg,Fischer:2006ub}.


\section{Vertices in three dimensions}\label{s3d}

\begin{figure*}
\includegraphics[width=\linewidth]{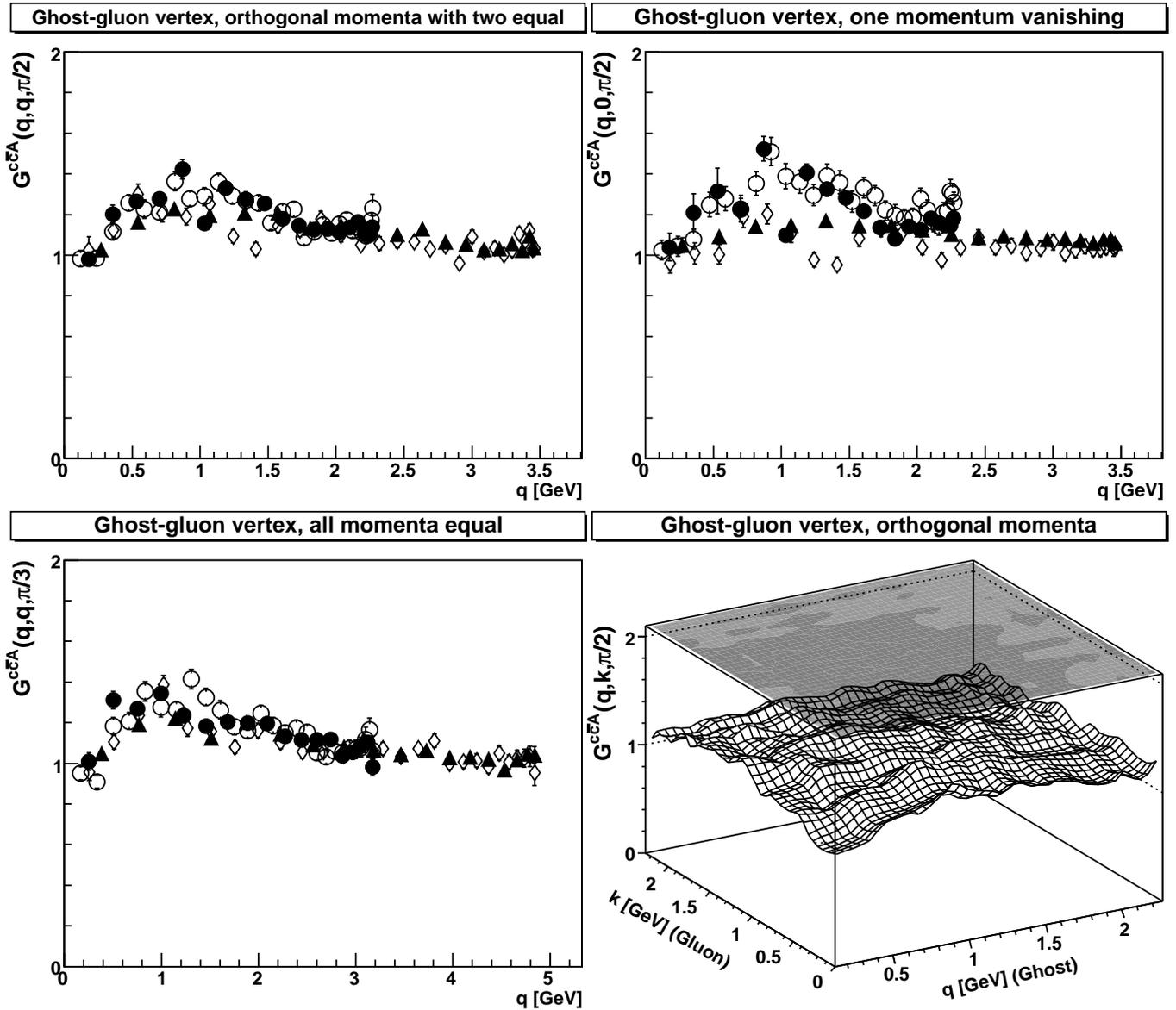}
\caption{\label{f3dggv}
The ghost-gluon vertex in three dimensions. The bottom-left panel shows the
so-called equal configuration. The bottom-right panel shows the orthogonal
configuration. The top panels give cuts through the orthogonal plane. The
top-left panel shows the case of ghost and gluon momentum with equal magnitude.
The top-right panel shows the case with a vanishing gluon momentum.
Full triangles are from a $40^3$ lattice at $\beta=6.0$,
open diamonds from a $60^3$ lattice at $\beta=6.0$,
full circles from a $40^3$ lattice at $\beta=4.2$
and open circles from a $60^3$ lattice at $\beta=4.2$.
The bottom-right panel shows the data only for the $60^3$ lattice at $\beta=4.2$.
All data are in physical units.}
\end{figure*}

\begin{figure*}
\includegraphics[width=\linewidth]{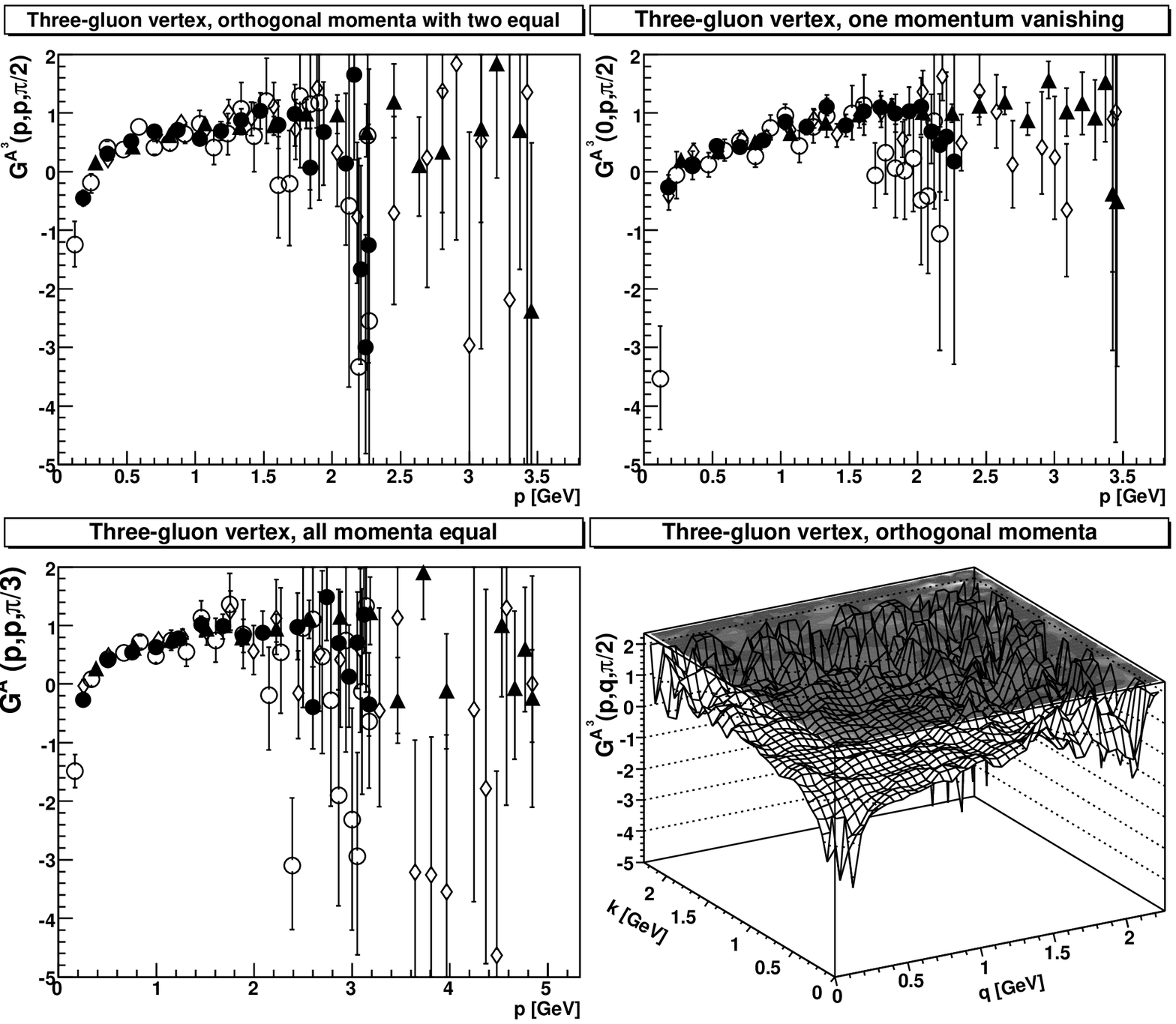}
\caption{\label{f3dg3v}
Same as in Fig.\ \ref{f3dggv} but the data now refer to the three-gluon vertex
in three dimensions.}
\end{figure*}

For the numerical determination of the vertices in three dimensions we
follow the same procedure used in the four-dimensional case. In particular, 
we consider again the contraction defined in Eq.\ \pref{proj} above.

The results for the ghost-gluon vertex, shown in Figure \ref{f3dggv}, are found
to be qualitatively similar to those obtained in 4d (see Section \ref{s4d} above
and Ref.\ \cite{gccvertex}). They are also in agreement with the results in 2d
\cite{Maas:2007uv} and with results obtained in 3d for smaller physical volumes
\cite{Cucchieri:2006tf}. In particular, the vertex is essentially flat and
constant in the infrared limit --- except for a small maximum at mid-momenta. As in
four dimensions, this result coincides with predictions obtained using functional
methods \cite{Schleifenbaum:2004id}. Note that the maximum of the vertex occurs
in the range $[0.5, 1]$ GeV, i.e.\ at slightly larger momenta than the maximum in
the gluon propagator \cite{Cucchieri:2003di}, and that the position of this
maximum agrees very well with analytic predictions \cite{Schleifenbaum:2004id}.

The results for the three-gluon vertex are shown in Figure \ref{f3dg3v}. Again,
the vertex is in qualitative agreement with results obtained in higher
\cite{Maas:2006qw} and in lower
dimensions \cite{Maas:2007uv}. In particular, it becomes negative at essentially
the same momentum where the maximum for the gluon propagator occurs. At even
smaller momenta, the quantity $G^{A^3}$ becomes rapidly large and negative,
suggesting a divergence, as in two dimensions \cite{Maas:2007uv}. As a consequence,
at least one of the two tensor structures contributing to $G^{A^3}$ should be infrared
divergent (with a negative pre-factor). Of course, if they both diverge, then (at least)
the term with the stronger divergence should have a negative pre-factor.
Let us recall that, in the case of one vanishing gluon momentum, only one of the
two tensor components (the one proportional to the tree-level component) contributes to
the three-gluon vertex \cite{Alles:1996ka}. Thus, in this case, the vertex self-energy
should be negative and larger than the tree-level result. Finally, let us note
that a divergence for the three-gluon vertex is predicted by functional
methods \cite{Alkofer:2004it,Huber:2007kc,Schleifenbaum:2006bq}, although the sign
of the the pre-factor is either not accessible \cite{Alkofer:2004it,Huber:2007kc}
or it is positive \cite{Schleifenbaum:2006bq}. Since the sign of this pre-factor depends on the
interplay between various contributions, this discrepancy will probably not be easily resolved.


\section{Summary}\label{ssum}

As shown above, we find that the ghost-gluon vertex is essentially constant for
all momenta in three and four dimensions. At the same time, the three-gluon vertex
is found to be suppressed at mid-momentum as well as at the smallest momentum reachable
in four dimensions. In three dimensions, a clear zero-crossing with a likely infrared
divergence is observed.

Combining these results with previous data in four \cite{gccvertex}, three
\cite{Cucchieri:2006tf} and in two dimensions \cite{Maas:2007uv}, it is suggestive that,
for any number of dimensions $d$ and for
the momentum configurations considered, the ghost-gluon vertex is infrared constant
and non-zero, while the three-gluon vertex is (negative) infrared divergent.
These results are in good agreement with predictions and assumptions in functional
calculations \cite{Alkofer:2004it,Huber:2007kc,Schleifenbaum:2004id,Schleifenbaum:2006bq}.

Let us also note that, in the case of the three-gluon vertex, the mid-momentum behavior
is quite different from the tree-level one. The consequences of this result for the
quark-gluon vertex, as well as the relevance for bound-state calculations,
are interesting open questions. Finally, it should be remarked that inspecting the
various terms in the Dyson-Schwinger equation of the gluon propagator makes it
clear that genuine two-loop contributions, usually neglected in such calculations, will
likely be important at intermediate momenta.


\acknowledgments
 
A.\ M. was supported by the DFG under grant numbers MA 3935/1-1 and MA 3935/1-2 and by the
FWF under grant number P20330.
A.\ C. and T.\ M. were partially supported by FAPESP (under grants \# 00/05047-5
and 05/59919-7) and by CNPq (including grant \# 476221/2006-4). The work of T.M. is
supported also by a fellowship from the Alexander von Humboldt Foundation.
The simulations for the volume $V = 48^4$ have been done on the IBM supercomputer
at S\~ao Paulo University (FAPESP grant \# 04/08928-3).
The ROOT framework \cite{Brun:1997pa} has been used in this project.


\appendix

\section{Technicalities}\label{atech}

\begin{table*}
\caption{\label{conf3}
Number of configurations considered in our numerical simulations.
The value of the lattice spacing $a$ has been taken from Ref.\ \cite{Cucchieri:2003di}
in three dimensions and from Ref.\ \cite{Bali} in four dimensions.
{\em Sweeps} indicates the number of sweeps between two consecutive gauge-fixed measurements.
More details on the generation of the configurations, error-determination, etc.\ can be found
in Ref.\ \cite{Cucchieri:2006tf}.}
\begin{ruledtabular}
\vspace{1mm}
\begin{tabular}{|c|c|c|c|c|c|c|c|}
$d$ & Vertex & $N$ & $\beta$ & $a^{-1}$ [GeV] & Configurations & Sweeps & $L = V^{1/d}$ [fm]\cr
\hline
3 & Ghost-gluon & 40 & 6.0 & 1.733 & 1267 & 50 & 4.5 \cr
\hline
3 & Ghost-gluon & 60 & 6.0 & 1.733 & 460 & 70 & 6.8 \cr
\hline
3 & Ghost-gluon & 40 & 4.2 & 1.136 & 1077 & 50 & 6.9 \cr
\hline
3 & Ghost-gluon & 60 & 4.2 & 1.136 & 367 & 70 & 10 \cr
\hline
3 & Three-gluon & 40 & 6.0 & 1.733 & 9709 & 50 & 4.5 \cr
\hline
3 & Three-gluon & 60 & 6.0 & 1.733 & 5017 & 70 & 6.8 \cr
\hline
3 & Three-gluon & 40 & 4.2 & 1.136 & 11095 & 50 & 6.9 \cr
\hline
3 & Three-gluon & 60 & 4.2 & 1.136 & 8114 & 70 & 10 \cr
\hline
4 & Ghost-gluon & 16 & 2.5 & 2.309 & 1336 & 30 & 1.4 \cr
\hline
4 & Ghost-gluon & 22 & 2.5 & 2.309 & 1248 & 50 & 1.9 \cr
\hline
4 & Ghost-gluon & 16 & 2.2 & 0.938 & 1351 & 30 & 3.4 \cr
\hline
4 & Ghost-gluon & 22 & 2.2 & 0.938 & 1043 & 50 & 4.7 \cr
\hline
4 & Ghost-gluon & 48 & 2.2 & 0.938 & 100 & 100 & 10.1 \cr
\hline
4 & Three-gluon & 16 & 2.5 & 2.309 & 11446 & 30 & 1.4 \cr
\hline
4 & Three-gluon & 22 & 2.5 & 2.309 & 6291 & 50 & 1.9 \cr
\hline
4 & Three-gluon & 16 & 2.2 & 0.938 & 8600 & 30 & 3.4 \cr
\hline
4 & Three-gluon & 22 & 2.2 & 0.938 & 5365 & 50 & 4.7 \cr
\hline
4 & Three-gluon & 48 & 2.2 & 0.938 & 3396 & 100 & 10.1 \cr
\end{tabular}
\end{ruledtabular}
\end{table*}

The lattice simulations have been performed essentially in the same way as
in our previous investigations on smaller lattices \cite{Cucchieri:2006tf}. In particular,
a standard Wilson action has been used. The parameters of the individual runs
can be found in Table \ref{conf3}. Gribov-Singer copy effects \cite{Gribov}
have not been taken into account.
Finally, the calculation of the vertices and the error determination has also been
performed as in Ref.\ \cite{Cucchieri:2006tf}. In particular, note that all errors
represent a 68\% confidence level. 


\begin{figure}
\includegraphics[width=\linewidth]{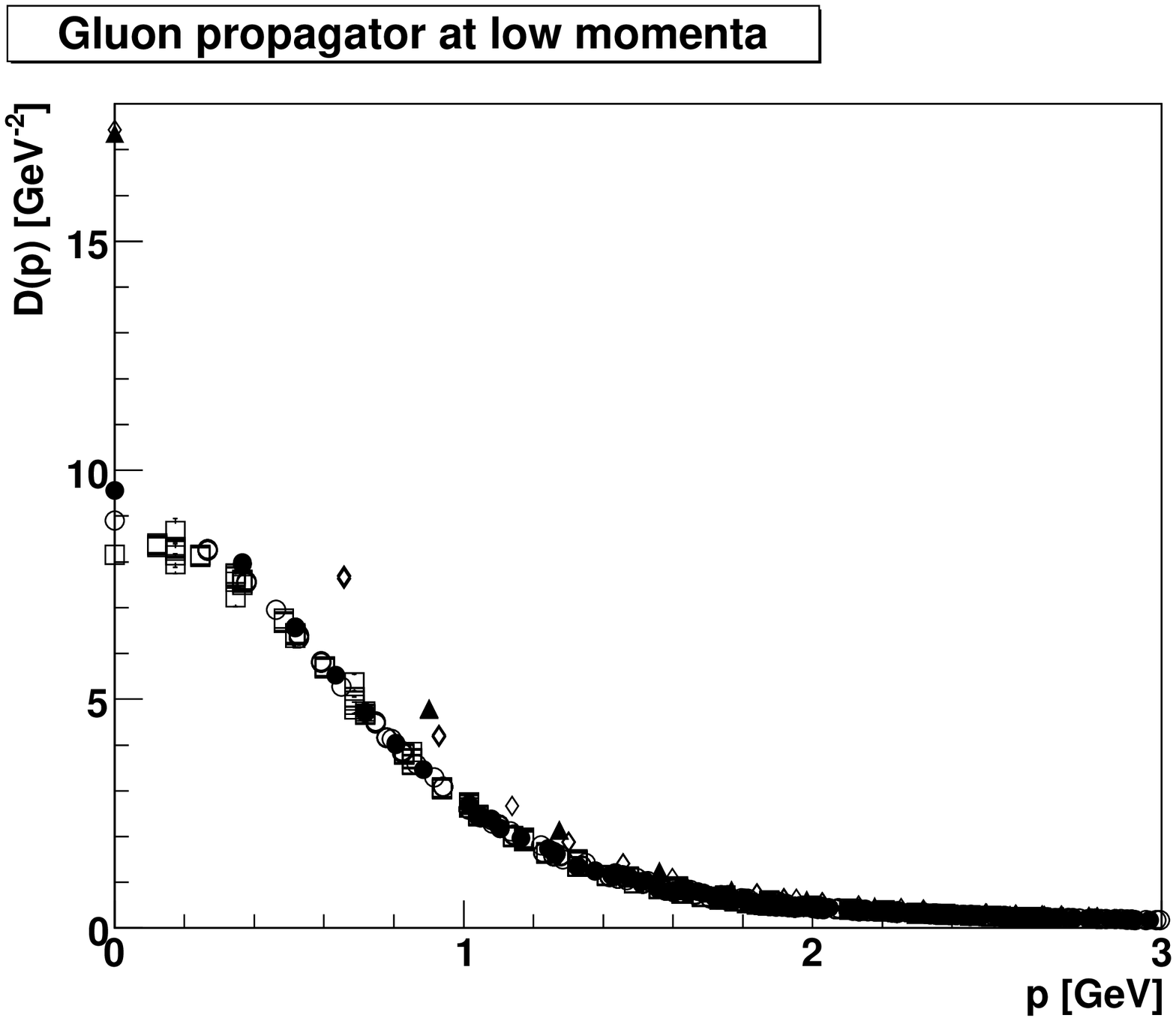}\\
\includegraphics[width=\linewidth]{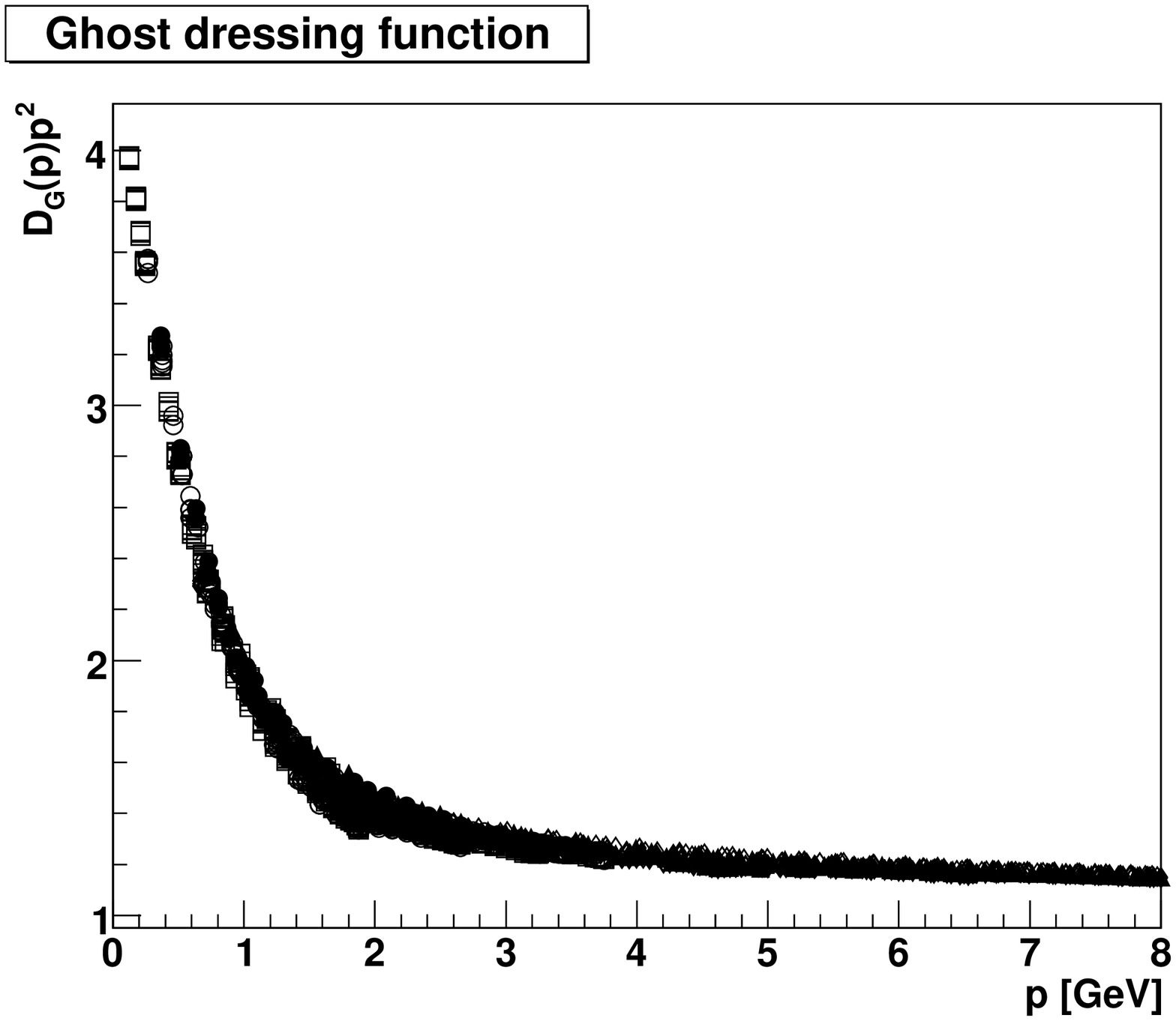}
\caption{\label{f4dprp}
Here we show the gluon propagator (top panel) and the ghost dressing function
(bottom panel), both in four dimensions. 
Full triangles are from a $16^4$ lattice at $\beta=2.5$,
open diamonds from a $22^4$ lattice at $\beta=2.5$,
full circles from a $16^4$ lattice at $\beta=2.2$,
open circles from a $22^4$ lattice at $\beta=2.2$
and open squares from a $48^4$ lattice at $\beta=2.2$.
}
\end{figure}

\begin{figure}
\includegraphics[width=\linewidth]{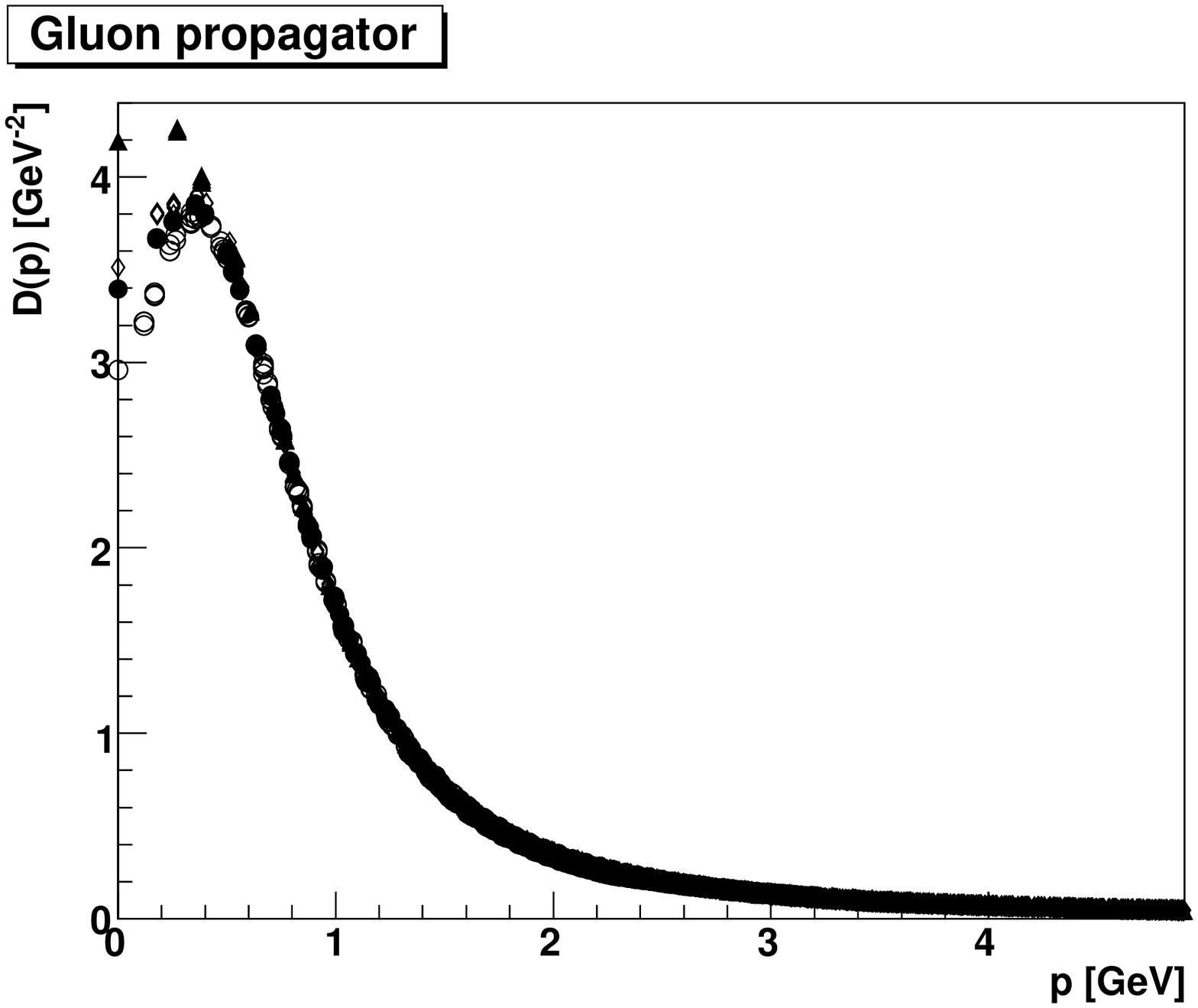}\\
\includegraphics[width=\linewidth]{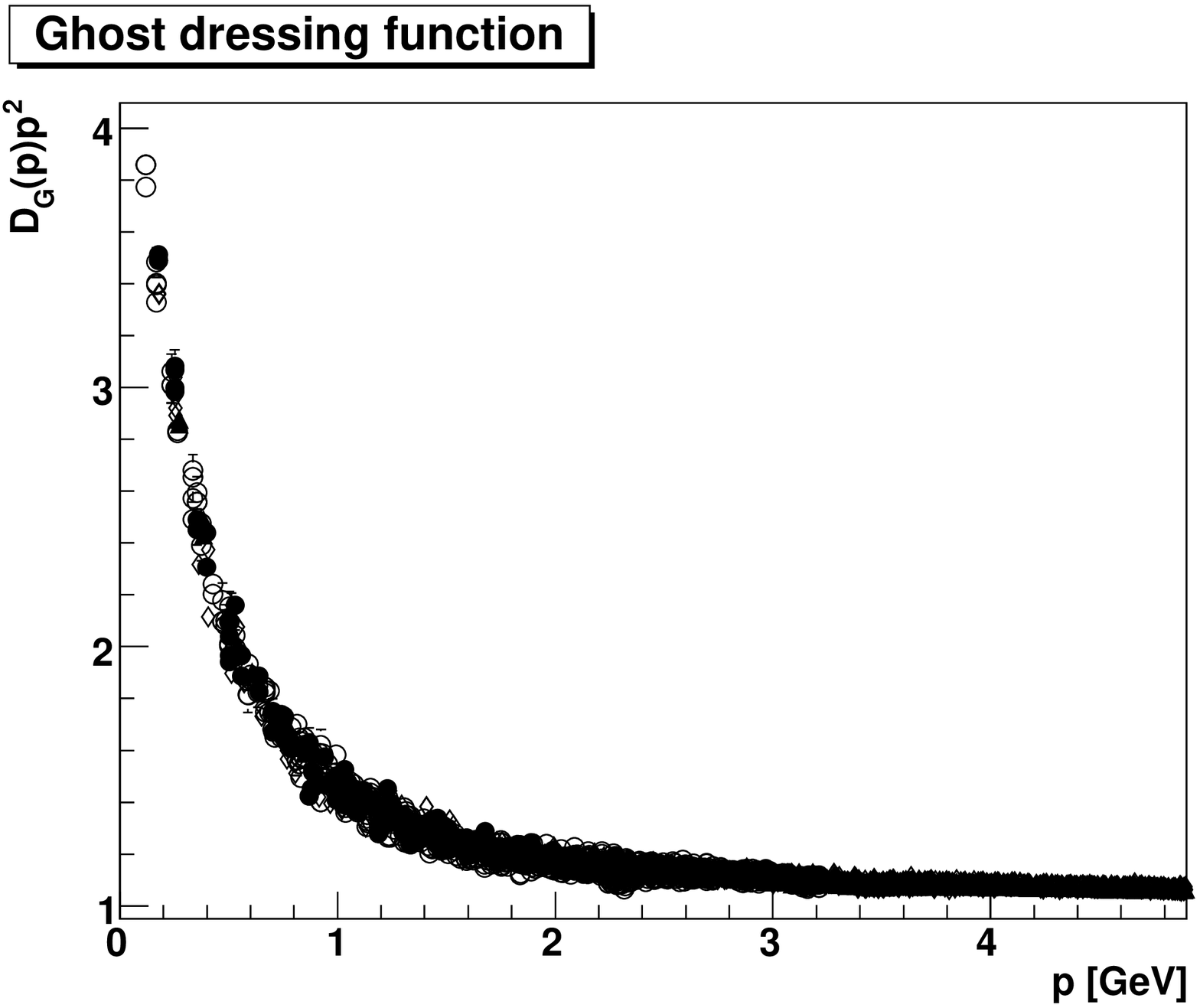}
\caption{\label{f3dprp}
Same as in Figure \ref{f4dprp} for the three-dimensional case.
Full triangles are from a $40^3$ lattice at $\beta=6.0$,
open diamonds from a $60^3$ lattice at $\beta=6.0$,
full circles from a $40^3$ lattice at $\beta=4.2$
and open circles from a $60^3$ lattice at $\beta=4.2$.}
\end{figure}

\section{Propagators}\label{aprop}

Since the gluon and ghost propagators are necessary in the process of amputating
the Green's functions in order to obtain the vertices, we report here (for
completeness) our data for the scalar part of the gluon propagator $D$ and for
the ghost propagator $D_G$. These data are obtained as described in Ref.\
\cite{Cucchieri:2006tf}. Results are shown in Figure \ref{f4dprp} for four dimensions
and in Figure \ref{f3dprp} for three dimensions.

As obtained in previous studies for similar lattice volumes, both in 3d and in 4d one
finds an infrared diverging ghost propagator. At the same time, when considering
the lattice volume $48^4$ at $\beta = 2.2$, the gluon propagator seems to display
a plateau or to get slightly suppressed at small momenta. On the other hand, there is a
distinct maximum in three dimensions for all but the smallest volume. Let us recall
that in the context of the Gribov-Zwanziger scenario
\cite{Gribov,Zwanziger:2003cf,gzwanziger} studies by Dyson-Schwinger equations predict
that all Green's functions behave in the far infrared like power laws
\cite{Alkofer:2004it,Fischer:2006vf} --- at least in the case when all momenta have the
same magnitude. These power laws have characteristic infrared exponents in the
continuum. Due to finite-volume effects, one expects that the effective infrared exponents
obtained from lattice simulations \cite{Cucchieri:2007ta,Fischer:2007pf} should converge
to the continuum results when the infinite-volume limit is taken. This is indeed the case
in two dimensions \cite{Maas:2007uv}. On the other hand, as said in the Introduction,
recent results in 3d and in 4d using very large lattices \cite{Oliveira:2007dy,cucchieril7,
Cucchieri:2007rg,limghost} show evidence that the gluon propagator is finite (and nonzero) at
zero momentum and that the ghost propagator has an infrared exponent very close to zero.



\end{document}